\documentclass[aps,prr,twocolumn,groupedaddress]{revtex4-1}

\usepackage[utf8]{inputenc}
\usepackage{graphicx}
\usepackage{bm}
\usepackage{amsmath}
\usepackage{amssymb}
\usepackage{siunitx}
\usepackage{mathtools}
\usepackage{hyperref}
\usepackage{physics}
\hypersetup{colorlinks=true,urlcolor=black,linkcolor=black,citecolor=black}

\newcommand\plot[3]{\begin{figure} \includegraphics[#2]{plot_#1}\caption{\label{fig:#1} #3}\end{figure}}

\newcommand\vek[1]{\vb{#1}}

\newcommand\ecom{\,,}
\newcommand\edot{\,.}

\newcommand{\gvek}[1]{\vb*{#1}}

\allowdisplaybreaks
\begin{document}

\title{Microscopic theory of magnetoconductivity at low magnetic fields in terms of Berry curvature and orbital magnetic moment}



\author{Viktor Könye}
\author{Masao Ogata}
\affiliation{Department of Physics, The University of Tokyo, Bunkyo-ku, Tokyo 113-0033, Japan}

\date{\today}
\begin{abstract}
Using a microscopic theory for the magnetoconductivity at low magnetic fields we show how the Hall and longitudinal conductivity can be calculated in the low scattering rate limit. In the lowest order of the scattering rate, we recover the result of the semiclassical Boltzmann transport theory. At higher order, we get corrections containing the Berry curvature and the orbital magnetic moment. We use this formalism to study the linear longitudinal magnetoconductivity in tilted Weyl semimetals. We discuss how our result is related to the semiclassical Boltzmann approach and show the differences that arise compared to previous studies related to the orbital magnetic moment.
\end{abstract}

\maketitle

\section{Introduction}

Electric transport in a magnetic field is an extensively studied topic of great importance in solid state physics with a long history. A widely used method to calculate conductivity is the semiclassical Boltzmann transport theory with relaxation time approximation  \cite{Boltzmann1872,Ziman2007,solyom2}, which is only valid in metallic states with definite Fermi surfaces. The Boltzmann transport theory can be further improved with the use of anomalous velocity coming from the Berry curvature \cite{Xiao2010,Chang1996,Sundaram1999}. At finite low magnetic fields the magnetoconductivity can be discussed with the Boltzmann theory \cite{Jones1934,Seitz1950,Pippard1989,Ziman2007}, and if the anomalous velocity is also included, the magnetoconductivity was shown to have a contribution coming from the Berry curvature
 \cite{Kim2014,Yip2015,Morimoto2016,Cortijo2016,Gao2017,Zyuzin2017,Nandy2018,Sun2019,Ma2019,Das2019,Xiao2020}. This gives rise to interesting phenomena such as the negative magnetoresistance caused by the chiral anomaly in Weyl semimetals
 \cite{Son2013,Li2016,Spivak2016,Andreev2018}. It was recently shown that in the linear order of the magnetic field we can get anomalous behavior such as the linear longitudinal magnetoconductivity
 \cite{Cortijo2016,Gao2017,Zyuzin2017,Ma2019,Das2019,Sun2019} and the linear planar Hall effect
 \cite{Nandy2018,Ma2019} in topological systems.

However, the Boltzmann transport theory is a semiclassical approximation, and it is not clear whether all the important contributions are included or not. Actually, it has been shown that, in the case of the orbital magnetic susceptibility, some coefficients of the contributions obtained in the Boltzmann theory with Berry curvature \cite{Gao2015} do not agree with the microscopically obtained results \cite{Hebborn1960,Ogata2017}. Furthermore, the Boltzmann transport theory cannot be applied, for example, to the cases with strong disorder where impurity bands play essential roles and there is no definite Fermi surface. Thus, it is necessary to formulate the magnetoconductivity in terms of a microscopic field theory, or in terms of Green’s functions and Kubo’s linear response theory. 

In the absence of a magnetic field Karplus and Luttinger 
\cite{Karplus1954} showed that a finite magnetic moment leads to an anomalous Hall conductivity that is expressed by what we call Berry curvature nowadays. It was later shown that the same result can be achieved with the semiclassical Boltzmann theory with anomalous velocity 
\cite{Jungwirth2002,Nagaosa2010}. In the case of no magnetic field the connection of the microscopic theory to the Boltzmann theory was discussed in Ref.
 \cite{Sinitsyn2007}.

Using linear response theory for finite magnetic fields is more challenging. At high magnetic fields the magnetoconductivity was studied by Abrikosov
 \cite{Abrikosov1998,Abrikosov2000}. His theory works at high fields when only the lowest Landau levels are occupied, and was successful in explaining the linear magnetoresistance observed in Dirac systems 
\cite{Kapitza1928,He2014}.

For small magnetic fields a microscopic theory for the Hall conductivity was developed by Fukuyama 
\cite{Fukuyama1969a,Fukuyama1969b}. In this theory the magnetoconductivity in the linear order of the magnetic field is given as a formula containing velocity operators and Green's functions. This formula was used to study the Hall conductivity of two dimensional massless fermions
 \cite{Fukuyama2007}. A similar approach was used to calculate the magnetoconductivity of graphene by Ando
 \cite{Ando2019}.

In this paper, we first obtain a formula equivalent to the Fukuyama’s formula \cite{Fukuyama1969a,Fukuyama1969b} but without the explicit bare mass of the electron. We also generalize it to the longitudinal magnetoconductivity, i.e., $\sigma_{zz}$ with the magnetic field parallel to the $z$-axis. It is shown that each case can be written in a one-line formula without the electron bare mass, which is very important when applied for various effective models. Another merit of this new formalism is that it is able to treat strong disorder, where the Boltzmann theory cannot be applied.

Then, in the weak-scattering limit we evaluate the formula in a general manner and show how this gives a normal contribution and an anomalous contribution connected to the Berry curvature and orbital magnetic moment, similarly to the no magnetic field case. In this approach the Berry curvature and the orbital magnetic moment appear naturally from the matrix elements of the velocity operator, without assuming an anomalous velocity connected to the Berry curvature or a Zeeman shift connected to the orbital magnetic moment.

To show the validity of the new formula, we study tilted Weyl semimetals as an example. We study both the weak and strong scattering cases and compare it to previous results \cite{Zyuzin2017,Ma2019,Das2019} that used the semiclassical Boltzmann approach. 

The formalism we discuss in this paper shares similarities with the calculation of orbital susceptibility. A well know method to calculate the orbital susceptibility is the Landau-Peierls formula
 \cite{Landau1930,Peierls1933}. This description is not complete, and a complete microscopic formalism was shown to contain the Landau-Peierls contribution with additional corrections \cite{Hebborn1960,Ogata2017,Hebborn1964,Fukuyama1969,Fukuyama1970,Fukuyama1971,Ogata2015,Raoux2015,Ogata2016,Ogata2016b}. This is analogous to our problem, but a big difference is that in the case of orbital susceptibility the scattering rate can be ignored, while in the case of magnetoconductivity it is essential to have finite results.

\section{Formalism}

We study systems that can be described with an effective Hamiltonian in the form of an $n\times n$ Bloch Hamiltonian $H_{\vek{k}}$. The eigenvalues and eigenvectors are denoted as $H_{\vek{k}}\ket{a,\vek{k}}=E_{a\vek{k}}\ket{a,\vek{k}}$.
For simplicity from now on $H_{\vek{k}}\equiv H$, $E_{a\vek{k}}\equiv E_a$, $\ket{a,\vek{k}}\equiv\ket{a}$, $\partial_{k_\mu}\equiv\partial_\mu$ and $\hbar=1$. The velocity operator is:
\begin{equation}
v_\mu=\partial_\mu H\edot
\end{equation}
The matrix elements of this can be expressed by derivating $\matrixel{a}{H}{b}=\delta_{ab}E_a$ \cite{Balasubramanian1990}:
\begin{equation}
\label{eq:currmatel}
v_{ab}^\mu=\delta_{ab}\partial_\mu E_a+(E_b-E_a)\braket{a}{\partial_\mu b}\edot
\end{equation}
We assume the Matsubara Green's function is diagonal in the eigenstate basis and can be expressed as:
\begin{align}
\label{eq:Green}
G_a(i\varepsilon_n)&=\frac{1}{i\varepsilon_n-E_a+\mu+i\Gamma_a(i\varepsilon_n)}\ecom
\end{align}
where $\Gamma_a(i\varepsilon_n)$ is the scattering rate, which describes the effects of the disorder.
The conductivity is calculated through the retarded current-current correlation function ($\Pi^R$) in the framework of linear response theory \cite{Bruus2004,Mahan1990}:
\begin{equation}
\sigma_{\mu\nu}=\lim\limits_{\omega\to 0}\frac{ie^2}{\omega}\Pi^R_{\mu\nu}(\omega)\ecom
\end{equation}
where $e>0$ is the elementary charge.  

In this paper we discuss the conductivity in a magnetic field up to linear order of the magnetic field. The uniform magnetic field points in the $z$ direction and we are interested in the Hall conductivity $\sigma_{xy}$ and longitudinal conductivity $\sigma_{zz}$. For the calculation of the current-current correlation we base our calculation on the microscopic theory of Hall conductivity developed by Fukuyama \cite{Fukuyama1969a,Fukuyama1969b}. Here the magnetic field is treated as a perturbation similarly to the electric field. In the zeroth order in the magnetic field, the current-current correlation can be expressed as:
\begin{align}
\label{eq:sig0}
\Pi_{\mu\nu}^{(0)}(i\omega_\lambda)=\frac{1}{\beta V}\sum\limits_{n,\vek{k}}&\mathrm{Tr}[v_\mu G_+v_\nu G]\ecom
\end{align}
where $G\equiv G(\vek{k},i\varepsilon_n)$ and $G_+\equiv G(\vek{k},i\varepsilon_n+i\omega_\lambda)$.
This order gives the usual conductivity \cite{Bruus2004,Mahan1990} and the anomalous Hall conductivity connected to the Berry curvature \cite{Karplus1954,Xiao2010,Cserti2010}.

In the first order of the magnetic field the current-current correlation function can be expressed as Eq. (\ref{eq:fukufull}) obtained by Fukuyama in Ref. \cite{Fukuyama1969b}. However, this formula contains the electron bare mass ($m$) and thus it is difficult to apply to effective models that do not have the electron bare mass. Here, we notice that the Fukuyama's formula can be rewritten as (for details on this see Appendix \ref{app:fukuyama}):
\begin{subequations}
\begin{align}
\nonumber
\label{eq:hallgen}
\Pi_{xy}^{(1)}(i\omega_\lambda)=-\frac{ieB}{\beta V}\sum\limits_{n,\vek{k}}&\mathrm{Tr}[v_x G_+v_y Gv_x Gv_y G-\\ &-v_x G_+v_y G_+v_x G_+v_y G]\ecom\\
\nonumber
\label{eq:longgen}
\Pi_{zz}^{(1)}(i\omega_\lambda)=-\frac{ieB}{\beta V}\sum\limits_{n,\vek{k}}&\mathrm{Tr}[v_z G_+v_z Gv_x Gv_y G-\\ &-v_z G_+v_y G_+v_x G_+v_z G]\ecom
\end{align}
\end{subequations}
which does not contain $m$. Therefore, the present new formalism can be applied to various effective models. Furthermore, we obtained a formula for $\Pi^{(1)}_{zz}$ which was not considered by Fukuyama, that gives a non-trivial longitudinal magnetoconductivity.

From now on we focus on these two terms that are proportional to the magnetic field.

\section{Magnetoconductivity}
The Eqs. (\ref{eq:hallgen}) and (\ref{eq:longgen}) formulas obtained for the magnetoconductivity can be applied for any strength of the scattering rate. In the following, we evaluate these formulas in the leading and subleading order, when the scattering rate is small, which is the situation where the Boltzmann theory is applied. 

We start with discussing the Hall conductivity. Using Eq. (\ref{eq:hallgen}) in the eigenstate basis the Hall conductivity can be expressed as:
\begin{subequations}
\label{eq:xywC}
\begin{align}
\sigma_{xy}^{(1)}&=-B\frac{e^3}{V}\sum\limits_{\vek{k}}\sum\limits_{a,b,c,d}v_{da}^xv_{ab}^yv_{bc}^xv_{cd}^y C_{abcd}\ecom\\
C_{abcd}&=-\lim\limits_{\omega\to0}\frac{1}{\beta\omega}\sum\limits_{n} G_a^+G_d(G_bG_c-G_b^+G_c^+)\ecom
\end{align}
\end{subequations}
where the $i\omega_\lambda=\omega+i\eta$ substitution was made and the $\eta\to0$ limit was taken in $C_{abcd}$. Using the Eq. (\ref{eq:currmatel}) form of the current operator we will have five different type of terms in Eq. (\ref{eq:xywC}) based on the number of Kronecker deltas. The detailed explanation on how the Matsubara summation can be evaluated is shown in Appendix \ref{app:hallcond}. The main assumption we use is that the scattering rate is constant $\Gamma_a(\varepsilon,\vek{k})\equiv\Gamma$ and small. We only keep terms proportional to $1/\Gamma^2$ ($\sigma_{xy}^{\text{norm}}$) and $1/\Gamma$ ($\sigma_{xy}^{\text{anom}}=\sigma_{xy}^{\text{berry}}+\sigma_{xy}^{\text{mag}}$) in the final expression. After several transformations the magnetic field dependent part of the Hall conductivity becomes ($\sigma_{xy}^{(1)}=\sigma_{xy}^{\text{norm}}+\sigma_{xy}^{\text{berry}}+\sigma_{xy}^{\text{mag}}$):
\begin{subequations}
\label{eq:sigxy}
\begin{align}
\nonumber
\sigma_{xy}^{\text{norm}}&=-\frac{Be^3\tau^2}{\hbar^4 V}\sum\limits_{\vek{k},a} f_a'\bigg\{\partial_xE_a\partial_{x}\partial_yE_a\partial_yE_a-\\
&\hspace{20pt}-\frac{1}{2}\Big[\left(\partial_xE_a\right)^2\partial_y^2E_a+\left(\partial_yE_a\right)^2\partial_x^2E_a\Big]\bigg\} \ecom\\
\sigma_{xy}^{\text{berry}}&=\frac{Be^3\tau\phantom{^2}}{\hbar^3 V}\sum\limits_{\vek{k},a}f_a'\partial_xE_a\partial_yE_a\Omega^a_{xy}\ecom\\
\nonumber
\sigma_{xy}^{\text{mag}}&=\frac{Be^3\tau\phantom{^2}}{\hbar^3 V}\sum\limits_{\vek{k},a}f_a'\bigg\{\partial_x\partial_yE_aM_{xy}^a-\\&\hspace{30pt}-\frac{1}{2}\Big[\partial_xE_a\partial_yM_{xy}^a+\partial_yE_a\partial_xM_{xy}^a\Big]\bigg\}\ecom
\end{align}
\end{subequations}
where $\Gamma=\hbar/2\tau$, $\Omega^a_{xy}$ is the $z$ component of the Berry curvature \cite{Berry1984,Xiao2010} and $M^a_{xy}$ is the $z$ component of the orbital magnetic moment \cite{Chang1996,Xiao2010}:
\begin{subequations}
\label{eq:berrorb}
\begin{align}
\Omega^a_{\mu\nu}&= i\Big[\braket{\partial_\mu a}{\partial_\nu a}-\mu\leftrightarrow \nu~\Big]\ecom\\
M_{\mu\nu}^a&=\frac{1}{2i}\Big[\matrixel{\partial_\mu a}{(E_a-H)}{\partial_\nu a}-\mu\leftrightarrow \nu~\Big]\edot
\end{align}
\end{subequations}

Quantities in Eq. (\ref{eq:berrorb}) can be expressed in an easier to evaluate formula where the derivatives of eigenstates do not appear. Using Eq. (\ref{eq:currmatel}) and the completeness of eigenstates:
\begin{subequations}
\label{eq:intrinsic}
\begin{align}
\Omega_{\mu\nu}^a&=i\sum\limits_{b\neq a}\frac{\matrixel{a}{\partial_\mu H}{b}\matrixel{b}{\partial_\nu H}{a}}{(E_a-E_b)^2}-\mu\leftrightarrow\nu\ecom\\
M_{\mu\nu}^a&=\frac{1}{2i}\sum\limits_{b\neq a}\frac{\matrixel{a}{\partial_\mu H}{b}\matrixel{b}{\partial_\nu H}{a}}{E_a-E_b}-\mu\leftrightarrow\nu\edot
\end{align}
\end{subequations}
These expressions are useful in the numerical calculations where an accurate derivative of the eigenstate with respect to momentum is difficult to obtain.

We continue with the longitudinal conductivity. Using Eq. (\ref{eq:longgen}) in the eigenstate basis the longitudinal conductivity can be expressed as:
\begin{subequations}
\begin{align}
\label{eq:zzwD}
\sigma_{zz}^{(1)}=-B\frac{e^3}{V}&\sum\limits_{\substack{\vek{k}\\a,b\\c,d}}v_{da}^zv_{ab}^z\left(v_{bc}^xv_{cd}^y D_{abcd}-v_{bc}^yv_{cd}^x \tilde{D}_{abcd}\right)\ecom\\
D_{abcd}&=-\lim\limits_{\omega\to0}\frac{1}{\beta\omega}\sum\limits_{n} G_a^+G_bG_cG_d\ecom\\
\tilde{D}_{abcd}&=-\lim\limits_{\omega\to0}\frac{1}{\beta\omega}\sum\limits_{n} G_aG_b^+G_c^+G_d^+\edot
\end{align}
\end{subequations}
Similarly to the Hall conductivity we can perform the Matsubara summations (for details see Appendix \ref{app:longcond}) and evaluate Eq. (\ref{eq:zzwD}) in the leading and subleading order with respect to $1/\Gamma\propto\tau$. We find that the longitudinal conductivity is expressed as ($\sigma_{zz}^{(1)}=\sigma_{zz}^{\text{norm}}+\sigma_{zz}^{\text{berry}}+\sigma_{zz}^{\text{mag}}$):
\begin{subequations}
\label{eq:sigzz}
\begin{align}
\sigma_{zz}^{\text{norm}}&=0 \ecom\\
\label{eq:sigzzQ}
\sigma_{zz}^{\text{berry}}&=-\frac{Be^3\tau\phantom{^2}}{\hbar^3 V}\sum\limits_{\vek{k},a}f_a'\partial_zE_a\left(2\grad E_a\cdot\vek{\Omega}^a-\partial_zE_a\Omega_z^a\right)\ecom\\
\sigma_{zz}^{\text{mag}}&= -\frac{Be^3\tau\phantom{^2}}{\hbar^3 V}\sum\limits_{\vek{k},a}f_a'\left(\partial_zE_a\partial_zM_z^a - \partial_z^2E_a M^a_z\right)\ecom
\end{align}
\end{subequations}
where $\Omega^a_\mu=\frac{1}{2}\varepsilon_{\mu\nu\eta}\Omega^a_{\nu\eta}$ and $M^a_\mu=\frac{1}{2}\varepsilon_{\mu\nu\eta}M^a_{\nu\eta}$.

The $\sigma_{xy}^{\text{norm}}$ and $\sigma_{zz}^{\text{norm}}$ terms are the normal contributions to the magnetoconductivity. These terms are the same as the ones obtained using the semiclassical Boltzmann transport theory with relaxation time approximation without anomalous velocity \cite{Jones1934,Seitz1950,Fukuyama1969b,Pippard1989,Ziman2007}. The terms proportional to $\tau$ are the anomalous contributions to the magnetoconductivity ($\sigma_{\mu\nu}^{\text{anom}}=\sigma_{\mu\nu}^{\text{berry}}+\sigma_{\mu\nu}^{\text{mag}}$).  The $\sigma_{xy}^{\text{berry}}$ and $\sigma_{zz}^{\text{berry}}$ terms contain the Berry curvature. Introducing the anomalous velocity in the semiclassical Boltzmann theory these same terms were obtained in Refs. \cite{Yip2015,Cortijo2016,Nandy2018,Sun2019,Ma2019,Das2019}. The $\sigma_{xy}^{\text{mag}}$ and $\sigma_{zz}^{\text{mag}}$ terms contain the orbital magnetic moment. The effect of the orbital magnetic moment is captured in the semiclassical theories with the energy correction caused by the magnetic field coupling to the orbital magnetic moment \cite{Xiao2010,Cortijo2016,Nandy2018,Sun2019}. The term containing the derivative of the orbital magnetic moment in $\sigma_{zz}^{\text{mag}}$ was obtained explicitly in Ref. \cite{Cortijo2016}. However, the terms proportional to the orbital magnetic moment are absent from these theories.

Note that if the system is time reversal symmetric $E_a(\vek{k})=E_a(-\vek{k})$, $\partial_\mu E_a(\vek{k})=-\partial_\mu E_a(-\vek{k})$, $\Omega_\mu^a(\vek{k})=-\Omega_\mu^a(-\vek{k})$ and $M_\mu^a(\vek{k})=-M_\mu^a(-\vek{k})$. These relations guarantee that $\sigma_{xy}^{\text{berry}}=0$, $\sigma_{xy}^{\text{mag}}=0$, $\sigma_{zz}^{\text{berry}}=0$ and $\sigma_{zz}^{\text{mag}}=0$. Thus, in order to see an anomalous magnetoconductivity we need to break time reversal symmetry. This is consistent with the Onsager relations that prohibit the appearance of these terms if time reversal symmetry holds \cite{Pippard1989,Ziman2007}.

\section{Tilted Weyl node}
To show the validity of the new formula, we study the magnetoconductivity of a tilted Weyl node, and compare our results with those obtained using the semiclassical Boltzmann theory. We start with a general two-level system with the following Hamiltonian:
\begin{align}
H&=\vek{h}(\vek{k})\cdot\gvek{\sigma}+h_0(\vek{k})\sigma_0\ecom & E_\pm&=h_0\pm h\ecom
\end{align}
where $\sigma_\alpha$ are the Pauli matrices. Using Eq. (\ref{eq:intrinsic}) for the Berry curvature and orbital magnetic moment we get:
\begin{subequations}
\label{eq:berorb}
\begin{align}
\Omega_{\mu\nu}^\pm&=\mp\frac{1}{2}\frac{\vek{h}\cdot\left(\partial_\mu\vek{h}\cross\partial_\nu\vek{h}\right)}{h^3}\ecom \\
M_{\mu\nu}^\pm&=\frac{1}{2}\frac{\vek{h}\cdot\left(\partial_\mu\vek{h}\cross\partial_\nu\vek{h}\right)}{h^2}\edot
\end{align}
\end{subequations}

First, we discuss a single Weyl node without tilting described by the Weyl Hamiltonian:
\begin{align}
\vek{h}&=v\hbar\vek{k}\ecom & h_0&=0\ecom & E_\pm&=\pm v\hbar k\edot
\end{align}
The Berry curvature and orbital magnetic moment using Eq. (\ref{eq:berorb}) is:
\begin{align}
\gvek{\Omega}^\pm&=\mp\frac{1}{2}\frac{\vek{k}}{k^3}\ecom &
\vek{M}^\pm&=\frac{1}{2}\frac{\vek{k}}{k^2}\edot
\end{align}
The different components of the magnetoconductivity calculated from Eqs. (\ref{eq:sigxy}) and (\ref{eq:sigzz}) at zero temperature are:
\begin{subequations}
\label{eq:Weylcond}
\begin{align}
\sigma_{xy}^{\text{norm}}&=-\frac{v\mu Be^3\tau^2}{6\pi^2\hbar^3}\ecom & \sigma_{xy}^{\text{berry}}&=0\ecom &\sigma_{xy}^{\text{mag}}&=0\ecom \\
\sigma_{zz}^{\text{norm}}&=0\ecom & \sigma_{zz}^{\text{berry}}&=0\ecom &\sigma_{zz}^{\text{mag}}&=0\edot
\end{align}
\end{subequations}
Even though the Berry curvature and orbital magnetic moment are not vanishing, after integration the anomalous contributions vanish because of the mirror symmetries of the system.

In order to get a finite anomalous magnetoconductivity we introduce a small tilting ($t<1$) in the $k_z$ direction:
\begin{align}
\vek{h}&=v\hbar\vek{k}\ecom & h_0&=v\hbar tk_z\ecom & E_\pm&=v\hbar (tk_z\pm k)\edot
\end{align}
The Berry curvature and orbital magnetic moment is unchanged, but the tilting breaks the mirror symmetry in the dispersion relation, and the components of the zero temperature magnetoconductivity calculated from Eqs. (\ref{eq:sigxy}) and (\ref{eq:sigzz}) become:
\begin{subequations}
\begin{align}
\label{eq:hallt}
\sigma_{xy}^{(1)}&=\sigma_h\frac{3\tanh^{-1}(t)-3t}{t^3}\ecom &\sigma_h&=-\frac{v\mu Be^3\tau^2}{6\pi^2\hbar^3}\ecom\\
\label{eq:longt}
\sigma_{zz}^{(1)}&=\sigma_l t\ecom & \sigma_l&=-\frac{vBe^3\tau}{4\pi^2\hbar^2}\ecom
\end{align}
\end{subequations}
\plot{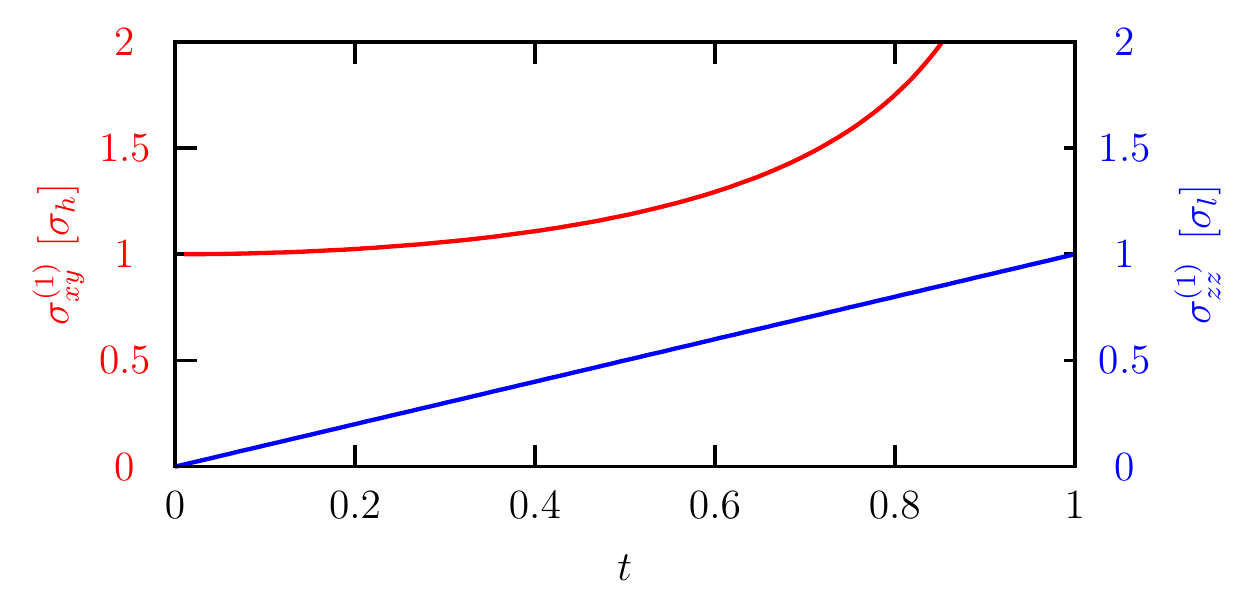}{width=.45\textwidth}{The Hall conductivity (\ref{eq:hallt}) and longitudinal magnetoconductivity (\ref{eq:longt}) of a tilted Weyl node with tilting $t$. $\sigma_h=-\frac{v\mu Be^3\tau^2}{6\pi^2\hbar^3}$ is the Hall conductivity at $t=0$ and $\sigma_l=-\frac{vBe^3\tau}{4\pi^2\hbar^2}$.}
where $\sigma_h$ is the Hall conductivity of the not tilted Weyl node in Eq. (\ref{eq:Weylcond}). The magnetoconductivity as a function of the tilting parameter is shown in Fig. \ref{fig:tilted}. As a consequence of the tilting a linear longitudinal magnetoconductivity appears.

This effect for the same model was also studied using the semiclassical Boltzmann transport theory in Refs. 
\cite{Zyuzin2017}, \cite{Ma2019} and \cite{Das2019}. The result obtained by Refs. \cite{Ma2019,Das2019} is:
\begin{align}
\label{eq:zzqsepsemi}
\sigma_{zz}^{\text{boltz}}&=\sigma_l\frac{-3t+5t^3+3t^5+3(t^2-1)^2\tanh^{-1}(t)}{3t^4}\edot
\end{align}
This is different from our result in Eq. (\ref{eq:longt}), but we find that it matches exactly the longitudinal magnetoconductivity calculated with only the Berry curvature contribution in Eq. (\ref{eq:sigzzQ}) ($\sigma_{zz}^{\text{boltz}}=\sigma_{zz}^{\text{berry}}$). This means that Refs. \cite{Ma2019,Das2019} failed to take into account the contribution coming from the orbital magnetic moment. In Fig. \ref{fig:tiltedsep} we show the longitudinal magnetoconductivity separated into contributions coming from the Berry curvature and orbital magnetic moment, and as we can see the terms containing the orbital magnetic moment significantly modify the result. The qualitative behavior of the result is not affected, but the quantitative value changes.
\plot{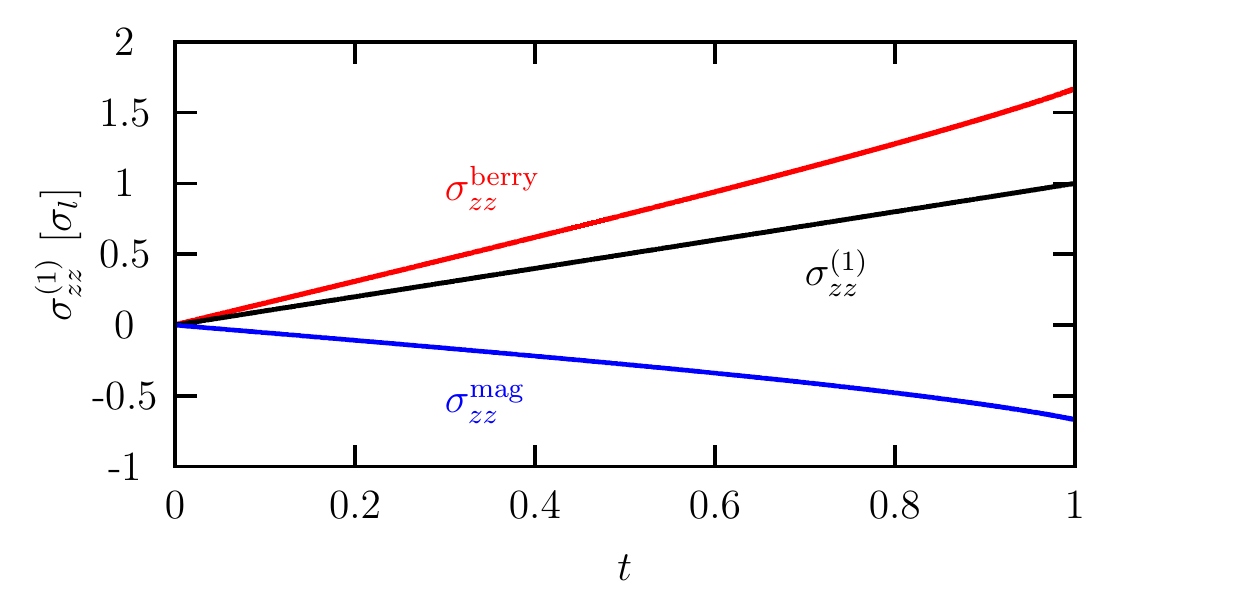}{width=.45\textwidth}{The longitudinal magnetoconductivity (\ref{eq:longt}) of a tilted Weyl node with tilting $t$. The red and blue line show the contribution of the Berry curvature and the orbital magnetic moment separately. $\sigma_l=-\frac{vBe^3\tau}{4\pi^2\hbar^2}$.}

In Ref. \cite{Zyuzin2017} the effect of the orbital magnetic moment is taken into consideration through the Zeeman shift of the energy. Their result, that the longitudinal conductivity is proportional to the tilting is consistent with our theory.

In real materials Weyl nodes come in pairs with opposite tilting and opposite chirality, and the Hamiltonian of the second Weyl node can be expressed as \cite{Zyuzin2017,Ma2019,Das2019}:
\begin{align}
\vek{h}&=-v\hbar\vek{k}\ecom & h_0&=-v\hbar tk_z\ecom & E_\pm&=-v\hbar (tk_z\pm k)\edot
\end{align}
Because of the sign change of both the tilting and the chirality the linear longitudinal conductivity persists even in the case of a pair of Weyl nodes. The total conductivity of two Weyl nodes will simply be twice of that of a single tilted Weyl node:

 \begin{equation}
         \sigma_{zz}^{2W}=2\sigma_{l}t\edot
 \end{equation}

This type of linear longitudinal magnetoconductivity has very unusual properties. The sign of the conductivity changes with the magnetic field, which can be used experimentally to distinguish this component from $\sigma_{zz}^{(0)}$ and $\sigma_{zz}^{(2)}$. It produces a negative magnetoresistance (or positive magnetoresistance, for opposite sign of the magnetic field) and gives an alternate mechanism to the chiral anomaly that produces a negative magnetoresistance. The chiral anomaly is also an effect that happens when the magnetic field and electric field are parallel, but the longitudinal magnetoconductivity in that case is quadratic in the magnetic field, and occurs without any tilting in the Weyl node. The symmetry properties in the magnetic field can be used to separate these two effects experimentally.

In order to estimate the magnitude of the linear longitudinal magnetoconductivity we can compare it to the conductivity at zero magnetic field and zero tilting calculated from Eq. (\ref{eq:sig0}):
\begin{align}
\sigma_{zz}^{(0)}=\frac{4\pi}{3}\frac{e^2}{h}\frac{\mu^2\tau}{h^2v}\edot
\end{align}
Since both quantities are proportional to the relaxation time, the ratio of $\sigma_{l}$ to $\sigma_{zz}^{(0)}$ becomes a $\tau$ independent number:
\begin{align}
\delta=\frac{\sigma_l}{\sigma_{zz}^{(0)}}=\frac{3}{2}\frac{\hbar v^2 e B}{\mu^2}\edot
\end{align}
Assuming realistic parameters such as $v=\SI{10e6}{\metre\per\second}$, $\mu=\SI{200}{\milli\electronvolt}$, and $B=\SI{1}{T}$ the ratio becomes $\delta\approx 0.025$. The effect is small, but not negligible and it can be enhanced with a smaller chemical potential.

\section{Large scattering rate}

In this section we discuss the large $\Gamma$ case. For this we have to go back to the Eq. (\ref{eq:zzwD}) form of the longitudinal conductivity. For the scattering rate in the Green's function (\ref{eq:Green}) we take
\begin{equation}
    \Gamma_a(\bm{k},\varepsilon)=\mathrm{sgn}(\mathrm{Im}(\varepsilon))\Gamma\ecom/
\end{equation}
where $\Gamma$ is constant. The Matsubara summation can be expressed as line integrals using the steps explained in the Supplementary material. At zero temperature these integrals can be evaluated analytically. The momentum integration is evaluated numerically. The longitudinal conductivity will be of the form:
\begin{align}
    \sigma_{zz}^{(1)} &= \sigma_l I(X,t)\ecom
\end{align}
where $X=\mu / \Gamma$ and $I$ is a dimensionless double integral in momentum space. We evaluated this double integral numerically and found that the longitudinal conductivity is independent of $X$. This means that $\sigma_{zz}=2\sigma_l t$ holds for any value of the scattering rate, which is unexpected in a simple Boltzmann theory.

\section{Discussion}

In this paper we studied the magnetoconductivity at low magnetic fields. We discussed the conductivity up to linear order of the magnetic field using linear response theory. We extended the microscopic formula for the Hall conductivity developed by Fukuyama 
\cite{Fukuyama1969a,Fukuyama1969b} to the longitudinal magnetoconductivity and evaluated it in a general manner for small scattering rates (Eqs. (\ref{eq:sigxy}) and (\ref{eq:sigzz})). These expressions were not derived before from a microscopic theory using a Green's function perturbative approach.

In the two lowest orders of the scattering rate we got terms of order $\order{\Gamma^{-2}}$ ($\sigma_{\mu\nu}^{\text{norm}}$) and $\order{\Gamma^{-1}}$ ($\sigma_{\mu\nu}^{\text{anom}}$). In the $\sigma_{\mu\nu}^{\text{norm}}$ term we recover the magnetoconductivity described by the semiclassical Boltzmann transport theory without anomalous velocity
\cite{Jones1934,Seitz1950,Fukuyama1969b,Pippard1989,Ziman2007}.

The components of order $\order{\Gamma^{-1}}$ are expressed in terms of the Berry curvature $\sigma_{\mu\nu}^{\text{berry}}$ and orbital magnetic moment $\sigma_{\mu\nu}^{\text{mag}}$. The $\sigma_{\mu\nu}^{\text{berry}}$ part was previously derived using the anomalous velocity in the Boltzmann theory 
\cite{Yip2015,Cortijo2016,Zyuzin2017,Nandy2018,Sun2019,Ma2019,Das2019}. The additional $\sigma_{\mu\nu}^{\text{mag}}$ term is present in Ref. 
\cite{Cortijo2016,Zyuzin2017} in a different form, only including the derivative of the orbital magnetic moment. In their theory the orbital magnetic moment appears through a Zeeman shift in the energy. In contrast, in our calculation the orbital magnetic moment appears naturally coming from the matrix element of the velocity operator.

The $\sigma_{\mu\nu}^{\text{berry}}$ and $\sigma_{\mu\nu}^{\text{mag}}$ quantities can only be nonzero if time reversal symmetry is broken. This is consistent with the Onsager relations, where these type of terms are forbidden if time reversal symmetry holds. An interesting symmetry property of the Hall conductivity is that the anomalous contribution is symmetric for the $x\leftrightarrow y$ change, while the normal contribution is antisymmetric.

Finally, we studied a tilted Weyl node using the above formalism. This system was discussed with the semiclassical Boltzmann theory \cite{Zyuzin2017,Ma2019,Das2019}. Our microscopic treatment is new in the literature. We showed that in a tilted Weyl node a finite linear longitudinal magnetoconductivity $\sigma_{zz}^{(1)}$ is present which is proportional to the tilting. This term was also found using the Boltzmann transport theory
 \cite{Ma2019,Das2019}, but only the effects of the Berry curvature were discussed. We discussed the effects of the orbital magnetic moment on the magnetoconductivity explicitly, and showed that it gives significant quantitative corrections, but do not affect the overall qualitative behavior. In Ref. \cite{Zyuzin2017} the orbital magnetic moment was introduced through the energy shift caused by the magnetic field coupled to the orbital magnetic moment. This type of treatment gives consistent result to our result, a longitudinal magnetoconductivity that is linear in both the magnetic field and the tilting.

 To study the effects of the scattering rate we calculated the longitudinal magnetoconductivity at finite $\Gamma$. We showed numerically that the large $\Gamma$ effects are negligible compared to the leading order of $\Gamma^{-1}$ even if $\mu\ll\Gamma$. Unlike in graphene \cite{Fukuyama2007} where this effect is relevant, in our case the lowest order approximation in the scattering rate is sufficient to get a good description.

A finite tilting is present in many Weyl semimetals \cite{Katayama2006,Kobayashi2008,Goerbig2008,Soluyanov2015}, making this effect relevant experimentally. In real materials the Weyl nodes come in pairs with opposite chirality, but they also tilt in opposite direction. As we showed this ensures that the effect of linear longitudinal magnetoconductivity persists even with two Weyl nodes. As we discussed the effect is small, but not negligible. The effect is enhanced for smaller charge carrier densities, larger tiltings, or larger magnetic fields. The conductivity changes sign with the magnetic field, making it possible to distinguish it from the zero field conductivity.

In this paper we only discussed the small scattering rate limit, where the impurities were taken into account through impurity Green's functions. In order to have a complete description the vertex correction should also be taken into account. The vertex correction in this context was discussed only for the free electron gas in Ref. 
\cite{Fukuyama1969a}. The treatment for general Bloch electrons is much more complicated, and for simplicity we neglected it similarly to Ref. 
\cite{Fukuyama1969b}. We expect the effects of the vertex correction to be quantitative and not qualitative, but the proper treatment of it is out of the scope of this study.

One of the merits of the present formalism is that we can discuss, for example, the cases where the electrons form an impurity band due to the strong disorder in which the semiclassical Boltzmann theory can not be applied. Although this kind of applications remain as future problems, the present paper gives a solid basis for further studies using a microscopic analysis.

\begin{acknowledgments}
We thank the very fruitful discussions with H. Matsuura and H. Maebashi. This work was supported by Grants-in-Aid for Scientific Research from the Japan Society for the Promotion of Science (Grant No. JP18H01162), and by JST-Mirai Program (Grant No. JPMJMI19A1).  
\end{acknowledgments}
\appendix
\section{Current-current correlation}
\label{app:fukuyama}
In this appendix we show how the formula for the current-current correlation in Eqs. (\ref{eq:hallgen}) and (\ref{eq:longgen}) is calculated using the formalism in Refs. \cite{Fukuyama1969a,Fukuyama1969b}. In the linear order of the vector potential, the current-current correlation is expressed as \cite{Fukuyama1969b}:
\begin{align}
\label{eq:fukufull}
\nonumber
\Pi_{\mu\nu}^{(1)}=-&\frac{ie^3}{2m\hbar}\frac{1}{\beta V}\sum\limits_{n,\vek{k},\alpha}\mathrm{Tr}[v_\mu G_+v_\alpha G_+G- \\
\nonumber
 &~~~~-  v_\mu G_+Gv_\alpha G](q_\alpha A^q_\nu-q_\nu A^q_\alpha)-\\
\nonumber
-&\frac{ie^3}{2\hbar^3}\frac{1}{\beta V}\sum\limits_{n,\vek{k},\alpha,\beta}(q_\alpha A^q_\beta-q_\beta A^q_\alpha)\times\\
\nonumber
&\times\bigg\{\Tr[v_\mu G_+v_\alpha G_+v_\nu G v_\beta G]+ \\
\nonumber
 &~~~~+\Tr[v_\mu G_+v_\nu Gv_\alpha G v_\beta G]+ \\
 &~~~~+\Tr[v_\mu G_+v_\alpha G_+v_\beta G_+ v_\nu G]\bigg\}\ecom
 \end{align}

where $m$ is the electron bare mass. The $q_\alpha A^q_\beta-q_\beta A^q_\alpha$ combination can be expressed as $\varepsilon_{\alpha\beta\gamma}B_\gamma^q$ since $\curl{\vek{A}}=\vek{B}$ and $\vek{A}(\vek{r})=-i\vek{A}^q\mathrm{e}^{i\vek{qr}}$. Using $\vek{B}=(0,0,B)$, $q_\alpha A^q_\beta-q_\beta A^q_\alpha=\varepsilon_{\alpha\beta z}B$. For the Green's functions and the velocity operators the following ward identity holds \cite{Fukuyama1969b}:
\begin{equation}
\partial_\mu G = G v_\mu G\edot
\end{equation}
Using this and partial integrations, for example, the following identity holds:
\begin{align}
\nonumber
\sum\limits_{\vek{k}}&v_xG_+v_yGv_xGv_yG=-\sum\limits_{\vek{k}}\bigg(\frac{\hbar^2}{m}v_xG_+Gv_xG+\\ &+v_xG_+v_yG_+v_yGv_xG+v_xG_+v_yGv_yGv_xG\bigg)\edot
\end{align}
Using other similar identities to this one, we get the Hall conductivity and longitudinal conductivity used in the main part of the paper as in Eqs. (\ref{eq:hallgen}) and (\ref{eq:longgen}). The advantage of this form is that the electron mass does not appear explicitly and it can be easily applied to effective Hamiltonians.

\section{Hall conductivity}
\label{app:hallcond}

Starting from Eq. (\ref{eq:xywC}) we show how the summation over the eigenstates and the Matsubara summations can be evaluated and how the Eq. (\ref{eq:sigxy}) form of the Hall conductivity is reached. Using the Eq. (\ref{eq:currmatel}) form of the current operator we have five type of terms in Eq. (\ref{eq:xywC}) based on the number of Kronecker deltas. The terms containing three Kronecker deltas will vanish after summation so we only have to consider the other four type of terms $\sigma_{xy}^{(1)}=\sigma_{xy}^O+\sigma_{xy}^I+\sigma_{xy}^{II}+\sigma_{xy}^{IV}\ecom$ where the indices represent the number of Kronecker deltas. After evaluating the sums over the Kronecker deltas and renaming indices we get:
\begin{subequations}
\begin{align}
\sigma_{xy}^{IV}=&-B\frac{e^3}{V}\sum\limits_{\vek{k},a} (\partial_xE_a)^2(\partial_yE_a)^2C_{aaaa}\ecom\\
\nonumber
\sigma_{xy}^{II}=&-B\frac{e^3}{V}\sum\limits_{\vek{k},a,b}(E_a-E_b)^2\times\\
\nonumber
&\times\bigg[\partial_xE_a\partial_xE_b\braket{\partial_ya}{b}\braket{b}{\partial_ya}C_{abba}+\\
\nonumber
&+\partial_yE_a\partial_yE_b\braket{\partial_xa}{b}\braket{b}{\partial_xa}C_{aabb}+\\
\nonumber
&+\partial_xE_a\partial_yE_a\braket{\partial_ya}{b}\braket{b}{\partial_xa}(C_{aaab}+C_{abaa})+\\
&+\partial_xE_a\partial_yE_a\braket{\partial_xa}{b}\braket{b}{\partial_ya}(C_{aaba}+C_{baaa})\bigg]\ecom\\
\nonumber
\sigma_{xy}^{I}=&-B\frac{e^3}{V}\sum\limits_{\vek{k},a,b,c}(E_a-E_b)(E_b-E_c)(E_c-E_a)\times\\
\nonumber
&\times\big[\partial_yE_a\braket{a}{\partial_x c}\braket{b}{\partial_x a}\braket{c}{\partial_y b}(C_{aacb}+C_{cbaa})-\\
&-\partial_xE_a\braket{c}{\partial_y a}\braket{a}{\partial_y b}\braket{b}{\partial_x c}(C_{abca}+C_{caab})\big]\ecom\\
\nonumber
\sigma_{xy}^O=&-B\frac{e^3}{V}\sum\limits_{\vek{k},a,b,c,d}(E_a-E_d)(E_b-E_a)(E_c-E_b)\times\\ &\times(E_d-E_c)\braket{d}{\partial_x a}\braket{a}{\partial_y b}\braket{b}{\partial_x c}\braket{c}{\partial_y d}C_{abcd}\edot
\end{align}
\end{subequations}

The next step is to evaluate the Matsubara summations. The details of these are discussed in the supplementary material. We are interested in the low impurity case so we assume that the scattering rate is constant $\Gamma_a(\varepsilon,\vek{k})\equiv\Gamma$ and small. Therefore, we neglect terms of $\order{\Gamma^0}$. It can be shown that $\sigma_{xy}^I=\order{\Gamma^0}$. Keeping only the terms $\order{\Gamma^{-2}}$ and $\order{\Gamma^{-1}}$, only $\sigma_{IV}$, $\sigma_{II}$, and $\sigma_{O}$ remain:
\begin{subequations}
\begin{align}
\sigma_{xy}^{IV}=&-\frac{Be^3}{4\Gamma^2V}\sum\limits_{\vek{k},a} (\partial_xE_a)^2(\partial_yE_a)^2 f_a''\ecom\\
\nonumber
\sigma_{xy}^{II}=&-\frac{3Be^3}{4\Gamma^2V}\sum\limits_{\vek{k},a}\partial_xE_a\partial_yE_a  f_a' \left(\Theta^a_{xy}+\Theta^a_{yx}\right)+\\
\nonumber
&+\frac{Be^3}{2\Gamma V}\sum\limits_{\vek{k},a}\partial_xE_a\partial_yE_a  f_a' \Omega^a_{xy}-\\
&-\frac{Be^3}{4\Gamma V}\sum\limits_{\vek{k},a}\partial_xE_a\partial_yE_a  f_a'' i\left(\Theta^a_{xy}-\Theta^a_{yx}\right)\ecom\\
\sigma_{xy}^O=&-\frac{Be^3}{2\Gamma V}\sum\limits_{\vek{k},a} f_a' i\left(\Theta^a_{xy}-\Theta^a_{yx}\right)\left(\Theta^a_{xy}+\Theta^a_{yx}\right)\ecom
\end{align}
\end{subequations}
where $f_a=(\exp{\beta(E_a-\mu)}+1)^{-1}$ is the Fermi-Dirac distribution and using the completeness of the eigenvectors ($\sum_b \dyad{b}{b}=1$):
\begin{subequations}
\begin{align}
\Theta^a_{xy}&= \sum\limits_b (E_a-E_b)\braket{\partial_xa}{b}\braket{b}{\partial_ya}\ecom\\
\Omega^a_{xy}&= i\big(\braket{\partial_xa}{\partial_ya}-x\leftrightarrow y\big)\edot
\end{align}
\end{subequations}
Here $\Omega^a_{xy}$ is the $z$ component of the Berry curvature \cite{Berry1984,Xiao2010}. The quantity $\Theta^a_{xy}$ can be transformed to:
\begin{align}
\Theta^a_{xy}&=\matrixel{\partial_xa}{\partial_yH-\partial_yE_a}{a}\edot
\end{align}
Using $\partial_x\partial_yH=0$ and the derivative of $\matrixel{a}{\partial_yH}{a}=\partial_yE_a$ we can show the following (similarly to Ref. \cite{Ogata2015}):
\begin{align}
\Theta^a_{xy}+\Theta^a_{yx}&=\partial_x\partial_yE_a\edot
\end{align}
The imaginary part of $\Theta_{xy}^a$ is the orbital magnetic moment \cite{Chang1996,Xiao2010}:
\begin{align}
M_{xy}^a=\frac{1}{2i}\left(\Theta_{xy}^a-\Theta_{xy}^a\right)=\Im{\matrixel{\partial_x a}{(E_a-H)}{\partial_y a}}\edot
\end{align}

Using partial integrations and separating terms proportional to $1/\Gamma^2$ and $1/\Gamma$ we reach the Eq. (\ref{eq:sigxy}) form of the Hall conductivity.

\section{Longitudinal conductivity}
\label{app:longcond}
Starting from Eq. (\ref{eq:longgen}) we show how the summation over the eigenstates and the Matsubara summations can be evaluated and how the Eq. (\ref{eq:sigzz}) form of the longitudinal conductivity is reached. The derivation is similar to that of the Hall conductivity in the previous appendix.

After summation over the Kronecker deltas we get:
\begin{subequations}
\label{eq:zzD}
\begin{align}
\sigma_{zz}^{IV}=&-B\frac{e^3}{V}\sum\limits_{\vek{k},a} (\partial_zE_a)^2\partial_xE_a\partial_yE_a (D_{aaaa}-\tilde{D}_{aaaa})\ecom\\
\nonumber
\sigma_{zz}^{II}=&-B\frac{e^3}{V}\sum\limits_{\vek{k},a,b}(E_a-E_b)^2\times\\
\nonumber
&\times\bigg[\partial_zE_a\partial_xE_b\braket{\partial_za}{b}\braket{b}{\partial_ya}D_{abba}+\\
\nonumber
&+\partial_zE_a\partial_yE_b\braket{\partial_xa}{b}\braket{b}{\partial_za}D_{aabb}+\\
\nonumber
&+\partial_xE_a\partial_yE_a\braket{\partial_za}{b}\braket{b}{\partial_za}D_{baaa}+\\
\nonumber
&+\partial_zE_a\partial_zE_a\braket{\partial_xa}{b}\braket{b}{\partial_ya}D_{aaba}+\\
\nonumber
&+\partial_xE_a\partial_zE_a\braket{\partial_ya}{b}\braket{b}{\partial_za}D_{aaab}+\\
\nonumber
&+\partial_yE_a\partial_zE_a\braket{\partial_za}{b}\braket{b}{\partial_xa}D_{abaa}-\\
&-(x\leftrightarrow y, D\leftrightarrow\tilde{D})\bigg]\ecom\\
\nonumber
\sigma_{zz}^{I}=&-B\frac{e^3}{V}\sum\limits_{\vek{k},a,b,c}(E_a-E_b)(E_b-E_c)(E_c-E_a)\times\\
\nonumber
&\times\bigg[\partial_yE_a\braket{a}{\partial_z c}\braket{b}{\partial_x a}\braket{c}{\partial_z b}D_{cbaa}-\\
\nonumber
&-\partial_xE_a\braket{a}{\partial_y b}\braket{b}{\partial_z c}\braket{c}{\partial_z a}D_{caab}+\\
\nonumber
&+\partial_zE_a\braket{a}{\partial_x c}\braket{b}{\partial_z a}\braket{c}{\partial_y b}D_{aacb}-\\
\nonumber
&-\partial_zE_a\braket{a}{\partial_z b}\braket{b}{\partial_x c}\braket{c}{\partial_y a}D_{abca}-\\
&-(x\leftrightarrow y, D\leftrightarrow\tilde{D})\bigg]\ecom\\
\nonumber
\sigma_{zz}^O=&-B\frac{e^3}{V}\sum\limits_{\vek{k},a,b,c,d}(E_a-E_d)(E_b-E_a)(E_c-E_b)\times\\
\nonumber
 &\times(E_d-E_c)\big[\braket{d}{\partial_z a}\braket{a}{\partial_z b}\braket{b}{\partial_x c}\braket{c}{\partial_y d}D_{abcd}-\\
&-(x\leftrightarrow y, D\leftrightarrow\tilde{D})\big]\ecom
\end{align}
\end{subequations}
The Matsubara summation is evaluated the same way as for the Hall conductivity (for details see the supplementary material) and we get:
\begin{subequations}
\label{eq:zzf}
\begin{align}
\sigma_{zz}^{IV}=&-B\frac{e^3}{4\Gamma^2V}\sum\limits_{\vek{k},a} (\partial_zE_a)^2\partial_xE_a\partial_yE_a f_a''\ecom\\
\nonumber
\sigma_{zz}^{II}=&-B\frac{e^3}{4\Gamma^2V}\sum\limits_{\vek{k},a}\partial_zE_a\big[\partial_xE_a\partial_y\partial_zE_a+\partial_yE_a\partial_z\partial_xE_a+\\
\nonumber
&~~~~~~~~~~~~~~~~~~~~+\partial_zE_a\partial_x\partial_yE_a\big]-\\
\nonumber
&-B\frac{e^3}{2\Gamma V}\sum\limits_{\vek{k},a}\partial_zE_a\Big\{f_a'\grad E_a\cdot\gvek{\Omega}^a+f_a''\grad E_a\cdot\vek{M}^a+\\
&+if_a'\sum\limits_b\left[\left(\partial_xE_b\braket{\partial_ya}{b}\braket{b}{\partial_za}-z\leftrightarrow y\right)-x\leftrightarrow y\right]\Big\}\ecom
\end{align}
\begin{align}
\nonumber
\sigma_{zz}^{I}=&-B\frac{ie^3}{2\Gamma V}\sum\limits_{\vek{k},a}f_a'\partial_zE_a\times\\
\nonumber
&\times\Bigg\{\partial_z\partial_xE_a\braket{a}{\partial_ya}-\partial_y\partial_zE_a\braket{a}{\partial_xa}-\\
\nonumber
&-\sum\limits_b\left[\left(\partial_xE_b\braket{\partial_ya}{b}\braket{b}{\partial_za}-z\leftrightarrow y\right)-x\leftrightarrow y\right]+\\
&+\bigg[\Big(\matrixel{\partial_y a}{\partial_x H}{\partial_z a}-y\leftrightarrow z\Big)-x\leftrightarrow y\bigg]\Bigg\}\ecom\\
\sigma_{zz}^O=&-B\frac{e^3}{2\Gamma V}\sum\limits_{\vek{k},a}f_a'\big[\partial_{z}\partial_{x}E_aM_x^a-\partial_{y}\partial_{z}E_aM_y^a\big]\ecom
\end{align}
\end{subequations}

Using partial integrations the longitudinal conductivity will also have terms proportional to $\tau^2$ and $\tau$ and it can be expressed as in Eq. (\ref{eq:sigzz}). One of the partial integrations is not trivial, so we show it schematically here:
\begin{align}
\nonumber
&\sum\limits_{\vek{k},a}\partial_zE_af_a''\grad E_a\cdot\vek{M}^a=\\
\nonumber
-&\sum\limits_{\vek{k},a}\partial_zE_af_a''\partial_z E_aM^a_z+2f_a'\partial_z(\partial_z E_aM^a_z)+\\
&+f_a'\partial_x(\partial_z E_aM^a_x)+f_a'\partial_y(\partial_z E_aM^a_y)\edot
\end{align}
Using equations like $\partial_{x}\partial_{z}[(E_a-H)\ket{a}]=0$ this can be transformed and used to cancel some of the terms in Eq. (\ref{eq:zzf}).

\bibliography{bibliography.bib}

\pagebreak
~
\pagebreak
\begin{widetext}
\begin{center}
\textbf{\large Supplementary material - Microscopic theory of magnetoconductivity at low magnetic fields in terms of Berry curvature and orbital magnetic moment}
\end{center}
\setcounter{equation}{0}
\setcounter{figure}{0}
\setcounter{table}{0}
\setcounter{page}{1}
\makeatletter
\renewcommand{\theequation}{S\arabic{equation}}
\renewcommand{\thefigure}{S\arabic{figure}}
\renewcommand{\bibnumfmt}[1]{[S#1]}
\renewcommand{\citenumfont}[1]{S#1}
\end{widetext}
\section{Matsubara summation with branch cuts}
\begin{figure}
     \includegraphics[width=.20\textwidth]{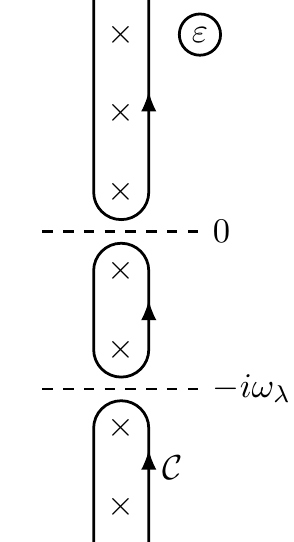}
     \includegraphics[width=.20\textwidth]{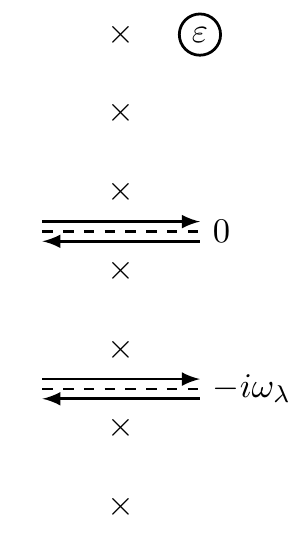}
     \caption{\label{fig:cont} The left side shows the contour integral equivalent to the Matsubara summation. The right side shows this same integral transformed to four ordinary integrals. The crosses show the singularities of the Fermi-Dirac distribution. The dashed lines show the branch cuts of $g$ in Eq. (\ref{eq:summI}).}
\end{figure}

The Matsubara summations in the main paper are all in the form of:
\begin{align}
\label{eq:summI}
I(i\omega_\lambda)&=\frac{1}{\beta}\sum\limits_n g(i\varepsilon_n,i\varepsilon_n+i\omega_\lambda)\ecom
\end{align}
where the function $g$ contains Green's functions with arguments $i\varepsilon_n$ or $i\varepsilon_n+i\omega_\lambda$. A simple example is:
\begin{equation}
g(i\varepsilon_n,i\varepsilon_n+i\omega_\lambda)=G_a(i\varepsilon_n)G_b(i\varepsilon_n+i\omega_\lambda)\edot
\end{equation}
In general there can be any number of Green's functions with any kind of indices. Because of the sign changing properties of the scattering rate at $\Im{\varepsilon}=0$, $g$ has two branch cuts and this type of summation can be transformed to four ordinary integrals using the residue theorem (for more details see Ref. \onlinecite{Bruus2004}):
    
\begin{align}
\nonumber
I^R(\omega)=-\int\limits_{-\infty}^\infty \frac{\dd{\varepsilon}}{2\pi i}f(\varepsilon)&\left[ g^{RR}(\varepsilon,\varepsilon+\omega)-g^{AR}(\varepsilon,\varepsilon+\omega)\right. +\\ &\left.+g^{AR}(\varepsilon-\omega,\varepsilon)-g^{AA}(\varepsilon-\omega,\varepsilon)\right]\ecom
\end{align}
where $f(\varepsilon)=(\exp{\beta(\varepsilon-\mu)}+1)^{-1}$ and we performed the analytic continuation in the frequency $i\omega_\lambda=\omega+i\eta$. The upper indices of $g$ show the retardedness of the Green's function with the corresponding argument. In our simple example:
\begin{align}
g^{XY}(\varepsilon,\varepsilon+\omega)&=G_a^X(\varepsilon)G_b^Y(\varepsilon+\omega)\\
G_a^{R/A}(\varepsilon)&=\frac{1}{\varepsilon-E_a\pm i\Gamma_a(\varepsilon)}\edot
\end{align}
The transformation of the Matsubara summation to integrals can be seen in Fig. \ref{fig:cont}

In the case of the conductivity only the $\omega\to0$ limit is important which can be expressed as:
\begin{align}
\label{eq:omlimI}
\nonumber
C=&-\lim\limits_{\omega\to 0}\frac{I^R(\omega)}{\omega}=\\
\nonumber
&=\int\limits_{-\infty}^\infty \frac{\dd{\varepsilon}}{2\pi i} f'(\varepsilon)\left[g^{AR}(\varepsilon,\varepsilon)-g^{AA}(\varepsilon,\varepsilon)\right]+\\&~~~~~+f(\varepsilon)\partial_\omega \left.\left[g^{RR}(\varepsilon,\varepsilon+\omega)-g^{AA}(\varepsilon,\varepsilon+\omega)\right]\right\vert_{\omega=0}\edot
\end{align}

\section{Integrals of Green's functions}
\begin{figure}
      \includegraphics[width=.40\textwidth]{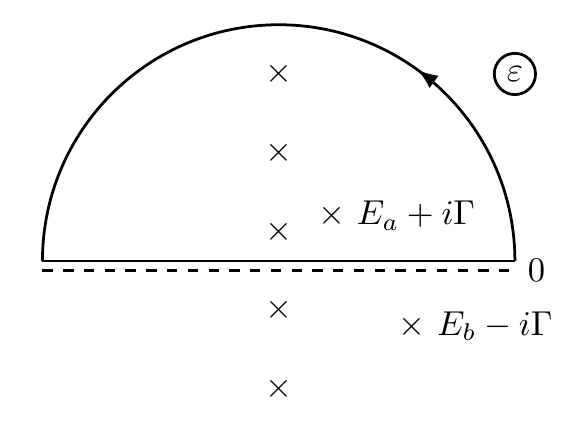}
      \caption{\label{fig:contupper}Contour integration on the upper half plane. The crosses show the singularities of the Fermi-Dirac distribution and the Green's functions.}
    \end{figure}

From now on we assume $\Gamma_a(\varepsilon)\equiv\Gamma$ and $\Gamma\to 0$. This means that we only keep the highest order terms in $\Gamma$ and neglect anything $\order{\Gamma^0}$. We substitute the infinite integral in Eq. (\ref{eq:omlimI}) with a contour integral on the upper complex plane as in Fig. \ref{fig:contupper}. The integrand will have several poles coming from the Fermi-Dirac distribution and poles coming from the advanced Green's functions inside the contour. After collecting the residues coming from the Fermi-distribution and performing the $\Gamma\to0$, limit we see that these contributions disappear since for $\Gamma=0$ the difference between advanced and retarded Green's functions disappears, thus in the combination $g^{AR}-g^{AA}$ and $g^{RR}-g^{AA}$ the singularities coming from the Fermi-Dirac distribution can be neglected in the order of $\order{\Gamma^0}$. This means that the integral can be substituted with the residues coming only from the advanced Green's functions in the upper plane. This same argument can be done with the lower half plane and retarded Green's functions, and the results do not change.

This means immediately that the term $g^{RR}$ can be neglected. It can also be shown that $g^{AA}$ can only have contributions of $\order{\Gamma^0}$. In order to have a higher order term two poles with the same energy but with different retardednesses are necessary. In this case in the residue a $1/(E_a-E_a+2i\Gamma)$ type of term appears which is of $\order{\Gamma^{-1}}$. With higher order poles higher orders of $\Gamma^{-1}$ can also appear. Since in $g^{AA}$ all the poles are on one side the contribution is $\order{\Gamma^0}$.

This way $C$ can be calculated as:
\begin{align}
C=\sum_i \Res{f'(\varepsilon)g^{AR}(\varepsilon,\varepsilon),\varepsilon_i}\ecom
\end{align}
where $\varepsilon_i$ are the singular points on the upper half plane of only $g^{AR}$. Taking our simple example in the case of $b=a$:
\begin{equation}
\label{eq:Ccalculator}
C=\Res{f'(\varepsilon)G^A_a(\varepsilon)G^R_a(\varepsilon),E_a+i\Gamma}=\frac{f'(E_a)}{2i\Gamma}\edot
\end{equation}
This is exactly the same result as using the usual $G^A_a(\varepsilon)G^R_a(\varepsilon)=\pi\delta(\varepsilon-E_a)/\Gamma$ approximation. But in cases where there are more Green's functions, this approximation can not always be used. Our method provides a systematic approach to evaluate these types of integrals. It is important to note that here $f'(E_a+i\Gamma)\approx f'(E_a)$ was used since we are neglecting terms of $\order{\Gamma^0}$. In cases with more Green's functions the Taylor expansion of $f$ is necessary to get a proper result as we will see in the following section.

\section{Summations in the main paper}
Here we detail a summation in the main paper and then list the rest. We start with $C_{aaaa}$. The $g^{AR}$ in this case is:
\begin{align}
g^{AR}=G_a^RG_a^A(G_a^AG_a^A-G_a^RG_a^R)\edot
\end{align}
using Eq. (\ref{eq:Ccalculator}):
\begin{align}
\nonumber
C_{aaaa}&=\frac{1}{2}\left(f'G^R_a\right)''\bigg\rvert_{\varepsilon=E_a+i\Gamma}-f'\left(G^R_a\right)^3\bigg\rvert_{\varepsilon=E_a+i\Gamma}=\\
\nonumber
&=-f''\left(G_a^R\right)^2\bigg\rvert_{\varepsilon=E_a+i\Gamma}+\frac{1}{2}f'''G^R\bigg\rvert_{\varepsilon=E_a+i\Gamma}=\\
\nonumber
&= \frac{f''(E_a+i\Gamma)}{4\Gamma^2}+\frac{f'''(E_a+i\Gamma)}{4i\Gamma}\approx\\
\nonumber
&\approx \frac{f''(E_a)}{4\Gamma^2}+\frac{f'''(E_a)i\Gamma}{4\Gamma^2}+\frac{f'''(E_a)}{4i\Gamma}=\\
&= \frac{f''(E_a)}{4\Gamma^2}\edot
\end{align}

All the other integrals can be done in a similar way. Here are the results for the $C$ summations in the main paper:
\begin{subequations}
\begin{align}
C_{aaaa}&=\frac{f_a''}{4\Gamma^2}+\order{\Gamma^0}\ecom\\
C_{abba}&=\order{\Gamma^0}\ecom\\
C_{aabb}&=\frac{1}{2i\Gamma}\frac{f_a'-f_b'}{(E_a-E_b)^2}+\order{\Gamma^0}\ecom\\
C_{abab}&=C_{aabb}\ecom\\
\nonumber
C_{aaab}&=\frac{1}{4\Gamma^2}\frac{f_a'}{E_a-E_b}+\frac{i}{2\Gamma}\frac{f_a'}{(E_a-E_b)^2}-\\&-\frac{i}{4\Gamma}\frac{f_a''}{E_a-E_b}+\order{\Gamma^0}\ecom\\
C_{abaa}&=\frac{1}{2\Gamma^2}\frac{f_a'}{E_a-E_b}+\order{\Gamma^0}\ecom\\
C_{aaba}&=C_{abaa}\ecom~~~
C_{baaa}=C_{aaab}^*\ecom\\
C_{aacb}&=\frac{f_a'}{2i\Gamma}\frac{1}{E_a-E_b}\frac{1}{E_a-E_c}+\order{\Gamma^0}\ecom\\
C_{aacb}&=C_{abac}=-C_{cbaa}=-C_{baca}\ecom\\
C_{abca}&=\order{\Gamma^0}\ecom~~~
C_{caab}=\order{\Gamma^0}\ecom\\
C_{abcd}&=\order{\Gamma^0}\edot
\end{align}
\end{subequations}

The calculation of $D$ is done in a similar way to that of $C$:
\begin{subequations}
\begin{align}
C_{aaaa}&=D_{aaaa}-\tilde{D}_{aaaa}\ecom\\
D_{abba}&=\frac{1}{2i\Gamma}\frac{f_a'}{(E_a-E_b)^2}+\order{\Gamma^0}\ecom\\
D_{abba}&=\tilde{D}_{abba}=D_{aabb}=\tilde{D}_{aabb}=D_{abab}=\tilde{D}_{abab}\\
D_{baaa}&=\order{\Gamma^0}\ecom~~~
\tilde{D}_{baaa}=\order{\Gamma^0}\ecom\\
\end{align}
\begin{align}
\nonumber
D_{aaab}&=\frac{1}{4\Gamma^2}\frac{f_a'}{E_a-E_b}+\frac{i}{2\Gamma}\frac{f_a'}{(E_a-E_b)^2}-\\&-\frac{i}{4\Gamma}\frac{f_a''}{E_a-E_b}+\order{\Gamma^0}\ecom\\
\tilde{D}_{aaba}&=-D_{aaba}^*\ecom\\
D_{aaab}&=D_{aaba}=D_{abaa}\ecom\\
\tilde{D}_{aaab}&=\tilde{D}_{aaba}=\tilde{D}_{abaa}\ecom
\end{align}
\begin{align}
D_{aacb}&=\frac{f_a'}{2i\Gamma}\frac{1}{E_a-E_b}\frac{1}{E_a-E_c}+\order{\Gamma^0}\ecom\\
D_{aacb}&=\tilde{D}_{aacb}=D_{abca}=\tilde{D}_{abca}=D_{abac}=\tilde{D}_{abac}\ecom\\
D_{cbaa}&=\order{\Gamma^0}\ecom~~~
\tilde{D}_{cbaa}=\order{\Gamma^0}\ecom\\
D_{caab}&=\order{\Gamma^0}\ecom~~~
\tilde{D}_{caab}=\order{\Gamma^0}\ecom\\
D_{baca}&=\order{\Gamma^0}\ecom~~~
\tilde{D}_{baca}=\order{\Gamma^0}\ecom\\
D_{abcd}&=\order{\Gamma^0}\ecom~~~
\tilde{D}_{abcd}=\order{\Gamma^0}\edot
\end{align}
\end{subequations}

\end{document}